\documentclass[pra,twocolumn,showpacs,preprintnumbers,amsmath,amssymb]{revtex4}

\usepackage{graphicx}
\usepackage{dcolumn}
\usepackage{bm}

\begin{document}
\draft
\newcommand{\lw}[1]{\smash{\lower2.ex\hbox{#1}}}

\title{Stripe Formation in Fermionic Atoms on 2-D Optical Lattice inside
a Box Trap: DMRG Studies for Repulsive Hubbard Model with Open Boundary
Condition} 

\author{M.~Machida} 
\email{machida.masahiko@jaea.go.jp}
\author{M.~Okumura}
\email{okumura.masahiko@jaea.go.jp}
\author{S.~Yamada}
\email{yamada.susumu@jaea.go.jp}

\affiliation{ CCSE, Japan Atomic Energy Agency, 6-9-3 Higashi-Ueno,
Taito-ku Tokyo 110-0015, Japan}
\affiliation{CREST(JST), 4-1-8 Honcho, Kawaguchi, Saitama 332-0012,
Japan}

\date{\today}

\begin{abstract} 

We suggest that box shape trap enables to observe intrinsic properties
of the repulsive Hubbard model in a fixed doping in contrast to the
conventional harmonic trap bringing about spatial variations of atom
density profiles.  
In order to predict atomic density profile under the box trap, we apply
the directly-extended density-matrix renormalization group method to
4-leg repulsive Hubbard model with the open boundary condition. 
Consequently, we find that stripe formation is universal in a low hole
doping range and the stripe sensitively changes its structure with
variations of $U/t$ and the doping rate.  
A remarkable change is that a stripe formed by a hole pair turns to one
by a bi-hole pair when entering a limited strong $U/t$ range. 
Furthermore, a systematic calculation reveals that the Hubbard model
shows a change from the stripe to the Friedel like oscillation with
increasing the doping rate. 
\end{abstract}
\pacs{71.10.Fd, 71.10.Pm, 74.20.Mn, 03.75.Ss}

\maketitle

Ultra-cold atomic Fermi-gas \cite{ColdAtomF,FinLatt} has attracted not
only atomic gas community but also several physicists studying
strongly-correlated electron systems.  
The reason is that the so-called ``optical lattice'' \cite{OptLatt}
formed by counter laser beams creates a periodical lattice potential
described by the tight-binding model and the ``Feshbach resonance''
enables to access to the Hubbard model with the repulsive on-site
interaction. 
Thus, a research goal in the optical lattice with the Feshbach tuning
is to directly observe several controversial issues due to strong
correlation as seen in High-$T_{\rm c}$ cuprate superconductors and
other metal oxides in controllable manners \cite{toolbox}. 

\begin{figure}
\includegraphics[scale=0.35]{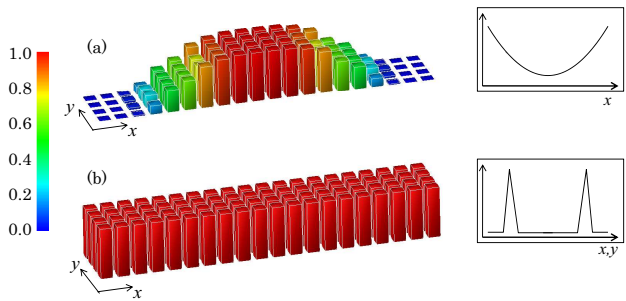}
\caption{\label{fig1} (a) A typical dex-DMRG calculation result for the
repulsive Hubbard model with a harmonic trap along the $x$ axis as shown
in the right-hand side panel. An open boundary condition is imposed
perpendicular to the $y$ axis. 
The lattice size is $20 \times 4$ (legs), and $U/t$ is 10. 
(b) A dex-DMRG result for the repulsive Hubbard model with the open
boundary condition for both axes. The trap shape on both axes is
schematically depicted in the right-hand side panel. The filling is just 
the half ($\sum_{\sigma} n_{\sigma} = 1$) and other conditions are the
same as (a).} 
\end{figure}

Generally, ultra-cold Bose and Fermi gases are trapped inside a harmonic
trap to avoid the free expansion of atoms.
The optical lattice is created inside the trap potential by utilizing
the field modulation in the standing waves of lasers.
Then, the model Hamiltonian on Fermi atoms (e.g., in 1-dimensional (1-D)
case) is described by the Hubbard model with a harmonic trap potential
\cite{1DTrap,1DTrapBoson} given by 
\begin{align}
H_{\rm Hubbard} & = -t \sum_{i,j,\sigma} ( a_{j \sigma}^\dag a_{i
 \sigma} + {\rm H.c.} ) + U \sum_i n_{i \uparrow} n_{i \downarrow}
 \nonumber \\ 
& \quad {} + V \left({\frac{2}{N-1}} \right)^2 \sum_{i,\sigma} n_{i
 \sigma} \left(i-\frac{N+1}{2}\right)^2 , \label{Hamiltonian}
\end{align}
where, a two-component Fermi gas is assumed, $a_{i\sigma}^\dagger$ is 
the creation operator of a Fermi atom with pseudo-spin $\sigma=\uparrow$
or $\downarrow$, $n_{i \sigma}(=a_{i\sigma}^\dagger a_{i\sigma})$ is the
density operator of the $i$-th site, $N$ is the total number of sites,
and the summation ($i,j$) in the first term describing the tunneling
between lattice site is usually taken over the nearest neighbor sites.
The on-site repulsive interaction $U$ is controllable via Feshbach
resonance, and the presence of the harmonic trap potential is
characterized by the last term including $V$. 
In the model Hamiltonian (\ref{Hamiltonian}), it is well-known that the 
atomic density profile gives spatial variation, whose typical one is
composed of the central Mott domain ($\sum_{\sigma} n_{i \sigma} \sim
1$) and the periphery metallic edges ($\sum_{\sigma} n_{i \sigma} < 1$) 
in a large repulsive $U/t$ range.
Such a variation is universal irrespective of the space dimension as
long as the harmonic well type of trap is employed. 
Actually, an example of 2-dimensional (2-D) case is shown in
Fig.~\ref{fig1} (a), in which the harmonic potential is applied only 
along the $x$ axis.
The result of Fig.~\ref{fig1} (a) is obtained in 4-leg Hubbard model by
using the directly-extended density-matrix renormalization group
(dex-DMRG) method \cite{Yamada,White,DMRGReview} to 2-D systems. 
See Ref.~\cite{1DTrap} for other characteristic variations of 1-D system 
in different parameter ranges. 

Such specific spatial-patterns have their own novel interests, e.g., how 
the spin correlation structures change from the Mott domain to metallic
edges and whether the holes in the metallic edges form Cooper pairs or
not \cite{Machida,Yamashita}. 
However, one clearly notices that intrinsic properties of the Hubbard
model under a fixed doping are not directly observable.
Although the system is quite clean and controllable, there is no direct
relationship with solid state physics except for a few artificial cases
\cite{Sigrist}. 
In this paper, we therefore suggest that an alternative trap, whose
shape is box (see the right hand side panel in Fig.~\ref{fig1} (b)), 
enables to experimentally study the Hubbard model with the open boundary
condition.  
A typical dex-DMRG result in the case is shown in Fig.~\ref{fig1} (b),
which displays atomic density profile in the half-filled case. This
result clearly shows that the Mott domain ($\sum_{\sigma} n_{i \sigma}
\sim 1$) extends over all region. 
By using such a trap shape, we can fully examine the Hubbard model under
a fixed doping and an interaction. 

\begin{figure}
\includegraphics[scale=0.25]{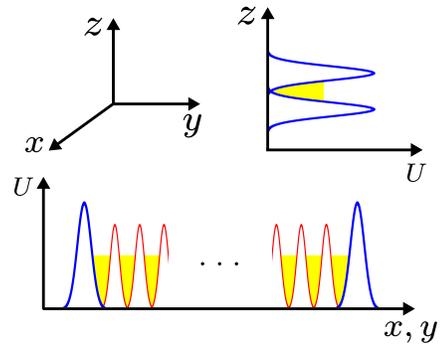}
\caption{\label{fig2} An expecting 2-D box-shape trap-potential with 
an optical lattice. The ``endcap beams'' \cite{Raizen} produce Gaussian
walls on the both ends of the $x$-, $y$-, and $z$-axis, while the
optical lattice is created by additional weak endcap beams between the
Gaussian walls.
The walls along the $z$ direction confines fermions much more tightly.}
\end{figure}

Recently, the 1-D box-shape confinement potential was actually created,
and the system was successfully condensed to Bose-Einstein condensate
inside the box \cite{Raizen}. 
According to Ref.~\cite{Raizen}, 1-D array of several small optical
wells is also available inside the box trap in a controllable manner. 
Thus, 1-D fermion Hubbard model with the open boundary condition is
accessible if bosons are replaced by fermions. 
Although the box trap dimension is now still one, the extension to 2-D 
box is expected to be straightforward. 
Our expecting potential shape is described in Fig.~\ref{fig2}, where 2-D 
box is created inside the $x$-$y$ plane with a narrow confinement along 
$z$-axis, and 2-D optical lattice is loaded inside the box plane by
adding the Gaussian walls \cite{Raizen} as shown in Fig.~\ref{fig2} or 
a standing wave light fields with shallow Gaussian envelope curve. 
Such a stage is just described by 2-D Hubbard model with the open
boundary condition. 
We emphasize that the system becomes the best playground for simulating 
controversial phenomena of High-$T_{\rm c}$ superconductors and other 
layered metal oxides. 
Furthermore, we would like to point out that ladder type models with the
open boundary are recently good targets for advanced DMRG methods
\cite{Yamada,WhiteStripe,Hager} and direct comparative studies are
possible. 
However, it should be noted that the box shape is just the first step to
remove the spatially dependent features of the filling. 
The temperature reduction in such a trap remains as the next step.   

A key issue in the repulsive 2-D Hubbard model with the open boundary is
the stripe formation, i.e., which types of stripes are formed with
variation of doping and $U/t$ still remains controversial.
Although the stripe formation have been also examined in $t$-$J$ model
\cite{t-JWorks}, we focus on solely the Hubbard model in this paper.
The reason is because $t$-$J$ model requires more specific conditions in
realizing its equivalent situation on atomic Fermi gases. 
Therefore, our research priority is the Hubbard model.
In addition, we note that the Hubbard model requires much more efforts
in numerically exploring the ground state than $t$-$J$ model since the
degree of freedom is much bigger than that of $t$-$J$ model \cite{nume}
and therefore the stripe formation and its profile in the ground state of
the Hubbard model is the present-day numerically challenging issue. 

Let us briefly review DMRG studies for the stripe formation of the
Hubbard model. 
After an early DMRG work on 3-leg Hubbard ladder \cite{Bonca}, White and
Scalapino applied the multichain algorithm of DMRG to 6-leg Hubbard
model and found the appearance of hole stripes \cite{WhiteStripe}.
Afterwards, Hager {\it et al.} re-examined it in a more systematic
manner \cite{Hager}. 
However, their results do not reach to the ground state because the
observed stripes still include the spin polarized modulation contrary to 
the Lieb-Mattis theorem as mentioned by themselves
\cite{Bonca,WhiteStripe,Hager}.
Although these pioneering works lacked final confirmations, they judged
the stripe states as the ground state from the convergent
tendencies. Thus, the accurate stripe profile in the ground-state are
still unknown and $U/t$ and filling dependence of the profile are also
unsolved. 
On the other hand, the authors have recently developed dex-DMRG and
confirmed that the hole stripe is really the ground state. 
The results converge and satisfy the Lieb-Mattis theorem.
This indicates that the present results give the first accurate profile
of the stripe in the Hubbard model. 
The extension to 2-D (ladder) is achieved by parallelizing the
diagonalization of the superblock Hamiltonian. 
See Ref.~\cite{Yamada} for its methodological details, performance, and
accuracy. 
This method gives us a chance of direct analysis on the ground-state. 
In this paper, we confine ourselves 4-leg ladder ($20 \times 4$)
Hubbard-model because the present dex-DMRG easily reaches its ground
state and enables to do systematic calculations varying $U/t$ and the
doping rate within our computational resources.
When using the present dex-DMRG, the 4-leg case $(20 \times 4)$ almost
converges within 3 times sweeps, which take about 3-hours under the use
of 128 CPU's on Altix 3700Bx2 in JAEA.
We obtain the ground state in a doping range from $p=0.025$ to $0.350$
with $U/t = 1 \sim 15$. 

\begin{figure}
\includegraphics[scale=0.36]{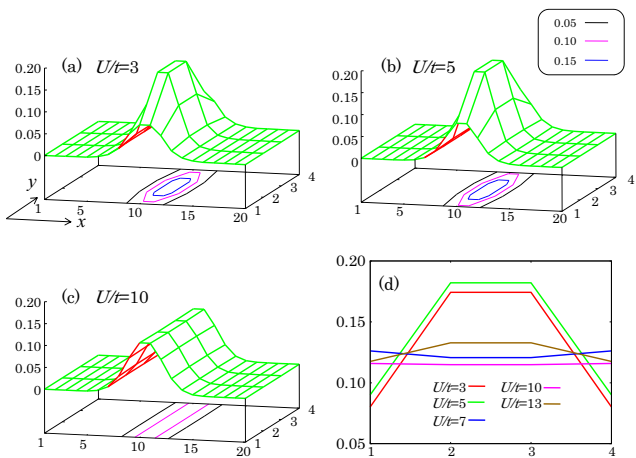}
\caption{\label{fig3} The dex-DMRG results of the hole distribution
profiles $1- (n_{i,\uparrow} + n_{i,\downarrow} )$ for $U/t = 3$ (a),
$5$ (b), and $10$ (c) in the Hubbard model with the open boundary
condition on the both axes. The doped hole is just a pair $(1 \!
\uparrow, \, 1 \! \downarrow)$ in all cases. The number of sites is $20 
\times 4$. (d) $U/t$ dependence of the hole distribution along the
ladder rung ($y$) direction at the center of the ladder leg $(x=10)$.}  
\end{figure}

Let us present numerical calculation results. 
At first, we focus on a case when a hole pair $(\uparrow, \, \downarrow)$
is doped into the half-filling. 
The doping rate is $p (\equiv 1-\frac{1}{N}\sum_{i,\sigma} n_{i,\sigma})
=0.025$, which corresponds to a heavily underdoped region in
High-$T_{\rm c}$ cuprate superconductors. 
The interest in the region is how the doped holes distribute, i.e.,
whether the two holes form a pair or not.  
If yes, then how the pair localizes (or delocalizes)?
Such a fundamental question is deeply relevant to unsolved issues in
heavily-underdoped High-$T_{\rm c}$ superconductors \cite{StripeReview}. 
An alternative experimental-stage is provided by the 2-D optical lattice
inside a 2-D box trap. 
Figure \ref{fig3} displays $U/t$ dependence of the hole distribution
profiles.  
In all cases, the spin-polarization completely drops to zero everywhere
according to the Lieb-Mattis theorem. 
One finds for all the three cases (a)--(c) that the doped two holes form
a stripe, called a hole-pair stripe. 
The area of the hole-pair stripe is much more compact than that of the
single hole ($\uparrow$ or $\downarrow$) doped case (not shown).
This means that an attractive interaction effectively works between two
holes.
In addition, the dex-DMRG results show that the stripe shape changes from
the oval to the complete stripe one with increasing $U/t$ (see 2-D
projected contour maps of (a) to (c)).
This means that the shape is affected by the boundary edge along the
ladder leg $(x)$ direction. 
As $U/t$ increases, the edge depression is considered to be defeated by
the energy cost-down due to the stripe-formation. 
Fig.~\ref{fig3} (d), which is $U/t$ dependence of the distribution
profile of the hole density along the ladder rung $(y)$ direction at the
center $(x=10)$, shows that there is a qualitative difference between
$U/t = 5$ and $7$ in the stripe shape. 
Next, in order to check the stability of the hole-pair stripe, we
examine its leg-length dependence and confirm that the stripe keeps 
the same shape up to $40\times4$ sites model. 
Although the maximum limit of the leg-length is presently $40$, any 
shape changes in the distribution profiles are not observable. 
Thus, one finds that two holes are bound and the pair sits in the most
stable location (the center) as a stripe. 

\begin{figure}
\includegraphics[scale=0.36]{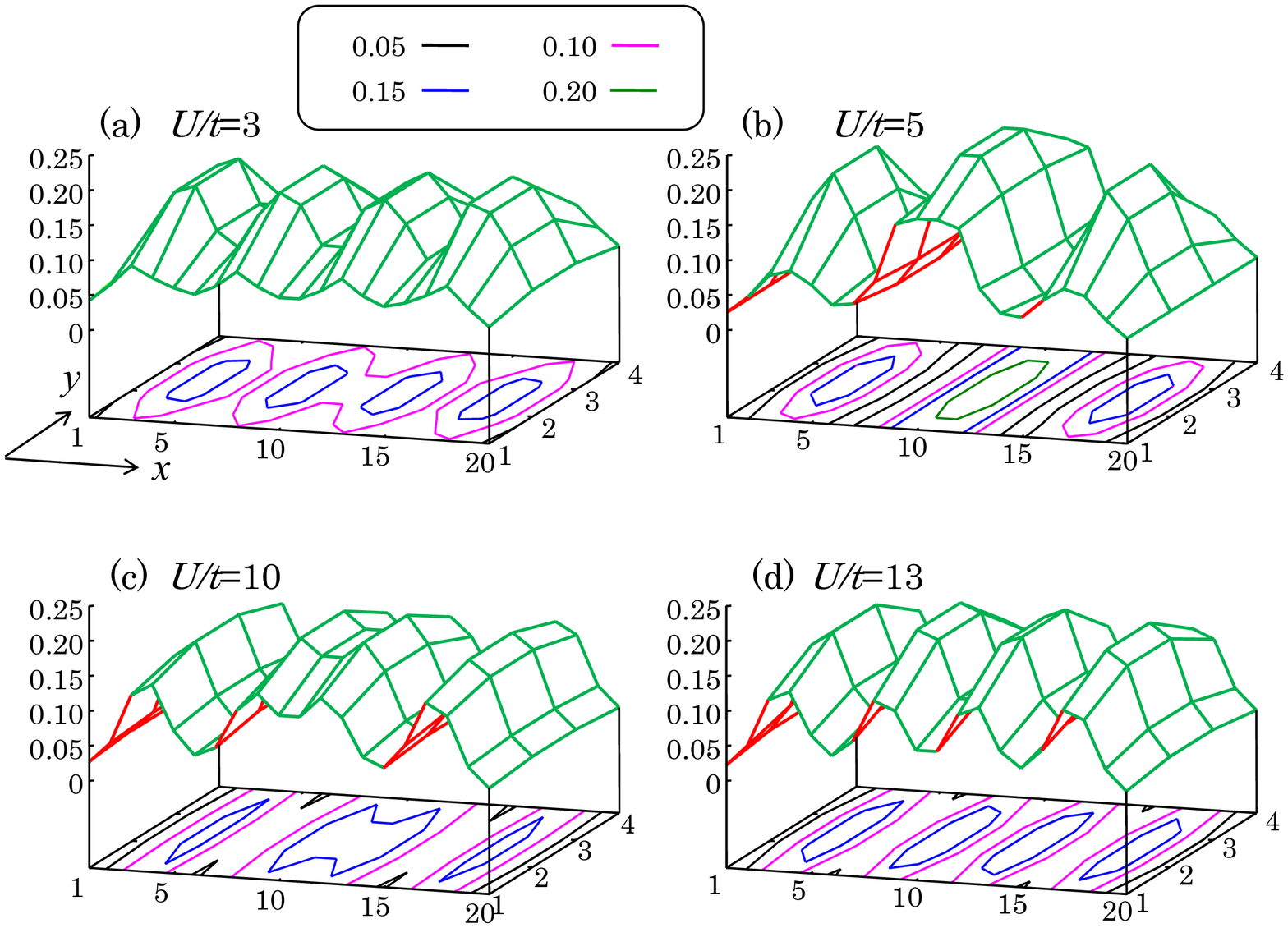}
\caption{\label{fig4} The dex-DMRG results of the hole density profiles
for (a) $U/t =3$ (b) $U/t = 5$, (b) $U/t =10$, and (d) $U/t= 13$ in the
Hubbard model with the open boundary condition. The number of doped
holes is 8 and the doping rate is $p=0.10$. The other conditions are
same as those in Fig.~\ref{fig3}.} 
\end{figure}

Next, let us study interaction between the hole-pair stripes.
The motivation comes from a question whether they simply merge or repel
each other as multiple hole-pair stripes closely stay. 
Such a situation shows up when holes are sufficiently doped.
If two hole-pair stripes merge, it means that the system prefers the
hole segregation \cite{StripeReview} or the stripe formation of a
bi-hole pair \cite{Chang}. 
On the other hand, if they are separated, then it indicates that the
hole-pair stripe is more stable. 
Here, we note that although such a study should include the size
dependence check, we concentrate on results only in a finite size system
($20\times 4$) due to lack of computational resources in this paper.
The size dependence will be reported elsewhere \cite{YamadaPrep}.
Figure \ref{fig4} shows $U/t$ dependence of the hole distribution
profile. 
The number of the doped holes is 8 and the doping rate is $p=0.10$,
which corresponds to an underdoped region in High-$T_{\rm c}$
superconductors \cite{StripeReview}. 
In Fig.~\ref{fig4} (a) ($U/t =3$), one finds 4 hole pair stripes, whose
distance is almost equivalent. 
It is found in this case that the stripes repel each other.
The surprising thing occurs at $U/t =5$, in which two stripes merge in
the central area and a large stripe appears. Moreover, the merged stripe
shows a stronger 1-D anisotropic feature (see and compare the 2-D
projected contour plots), because the hole density does not relatively
drop at the edges along the ladder leg $(x)$ direction. 
This stripe just corresponds to a bi-hole pair stripe, which extends over 
about 4 site.
Such a stripe structure characterized by $4$ holes was predicted by
Chang and Affleck \cite{Chang}. 
As $U/t$ increases further, the merged stripe starts to split into two
hole-pair stripes at $U/t = 10$, again, and the split becomes complete
at $U/t=13$. 
The bi-hole pair stripe is observable only for $5 \le U/t < 10$.
These behaviors indicate that the stripe formation occurs in not only a
simple manner but also multiple manners depending on the doping rate and
the interaction. 
The transition between these multiple structures may be too difficult to
clearly identify in solid state systems. Therefore, more controllable
optical lattice with the box shape trap is promising.
  
\begin{figure}
\includegraphics[scale=0.5]{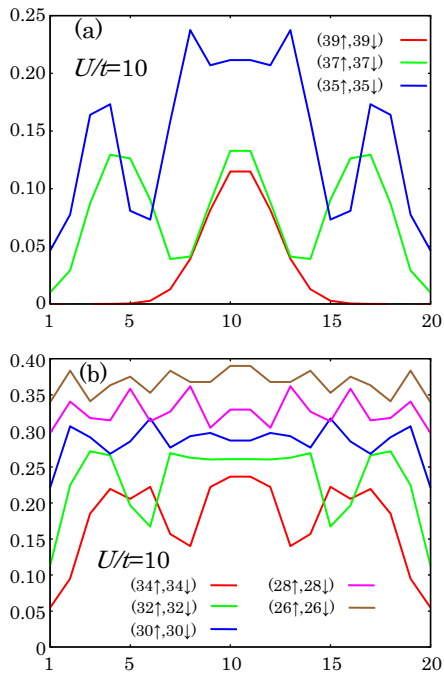}
\caption{\label{fig5} The dex-DMRG results of the hole density profile
along the ladder leg ($x$) direction of the 2nd(3rd)-leg for several
doping. $U/t$ is fixed to be 10. The other conditions are the same as
those in Fig.~\ref{fig3}} 
\end{figure}

Finally, let us show a systematic result of the doping-rate dependence
of the hole distribution profiles. The doping rate varies from $p=0.025$ 
to $0.350$, and $U/t$ is fixed to be $10$. In High-$T_{\rm c}$
superconductors, the heaviest doped rate ($p=0.350$) is in a overdoped
range, in which the materials almost show metallic behaviors. 
Figure \ref{fig5} shows a doping rate dependence on $x$-directional
profiles of hole distributions at $y=2$. 
It is found in Fig.~\ref{fig5} that there are two characteristic
profiles in the hole distribution. 
The first type seen in the relatively underdoped regime ($p \le 0.200(32
\! \uparrow \, 32 \! \downarrow $)) is characterized by the stripe,
whose modulation amplitude is relatively large (see Fig.~\ref{fig5} (a)
and (b)).  
In the regime, the stripes merge and the extended ones emerge as the
doped holes increases. 
The phenomenon is clearly observable from $p = 0.075$ $(37 \! \uparrow,
\, 37 \! \downarrow )$ to $p=0.200$ $(32 \! \uparrow, \, 32 \!
\downarrow )$. 
Inside this range, the bi-hole pair stripes and other stripes including
holes more than $4$ holes appear as well as the hole-pair ones.
Especially, in $p=0.125$ $(35 \! \uparrow, \, 35 \! \downarrow)$, we
point out that a bi-hole pair stripes develops quite distinctly. 
This doping rate is a characteristic one, where various singular
behaviors stemming from the stripe formation have been reported in 
High-$T_{\rm c}$ superconductors \cite{StripeReview}.  
In addition, one finds that two bi-hole pair stripes merge in $p=0.200$
$(32 \! \uparrow, \, 32 \! \downarrow)$ and a wider flat plateau emerges
compared to $p=0.150$ $(34 \! \uparrow, \, 34 \! \downarrow)$.
The second typical profile appears above $p=0.250$ $(30 \! \uparrow, \,
30 \! \downarrow)$. The spatial modulation structure in the hole profile
is independent on the doping rate.
The modulations in this metallic regime are Friedel like oscillations
\cite{FriedelOsc} 
rather than the stripe structures developed due to the strong
correlation \cite{Chang}. 

Before closing, let us discuss finite temperature effects on all the
present results. 
In atomic gases, the temperature reduction is always one of difficulties
in exploring the ground-state properties of many-body interacting
systems. 
The present results are static modulations of holes in the ground-state,
which are expected to fluctuate in non-zero temperature. 
However, in sufficiently low temperature, the resultant tiny fluctuation
does not matter for the static local-density probe.
On the other hand, in relatively high temperature, the static
modulations become fluctuating stripes, which can be observed by not
local but density correlation probes \cite{ColdReview}. 
Actually, dynamical stripe like fluctuations have been reported in
High-Tc superconductors \cite{StripeReview}, in which charge and spin
fluctuations are detectable much above the superconducting transition
temperature by various probes. 
Thus, the present stripe structures can be also identified by using
suitable correlation probes \cite{ColdReview} in finite temperature.

In conclusion, we suggested in atomic Fermi gases that the box shape
trap is essential for studies on intrinsic properties of the Hubbard
model with a fixed doping rate and an interaction. 
We applied the dex-DMRG method to the 4-leg Hubbard model with the open
boundary condition in order to predict rich varieties of hole profiles
including stripe formations due to strong correlation. 
Consequently, we observed the hole-pair stripe, the bi-hole pair stripe,
and other stripe structures including more than 4 holes in the
underdoped region. 
These structures merge and split with the variation of $U/t$ and the
doping rate.  
On the other hand, such characteristic features disappear and only a
monotonic oscillation pattern emerges in the overdoped one. 
These results can be directly confirmed in atomic Fermi gases, and
moreover, more systematic studies are possible in experiments of atomic
Fermi gases.
We believe that the optical lattice inside the box shape trap can solve
significant controversial issues like the stripe formation in
High-$T_{\rm c}$ superconductors and other metal oxides. 

Two of authors (S.Y. and M.M.) acknowledge M.~Kohno, T.~Hotta, and
H.~Onishi for illuminating discussion about the DMRG
techniques. M.M. also thanks Y.~Ohashi and H.~Matsumoto for the Hubbard 
model. 
The work was partially supported by Grant-in-Aid for Scientific Research
on Priority Area ''Physics of new quantum phases in superclean
materials'' (Grant No. 18043022) from the Ministry of Education,
Culture, Sports, Science and Technology of Japan. This work was also
supported by Grant-in-Aid for Scientific Research from MEXT, Japan
(Grant No.18500033), and a JAEA internal project represented by 
M.~Kato.


\begin{references}
\bibitem{nume} It should be noted that the degree of freedom of $t$-$J$
model is strongly restricted compared to the Hubbard model. Thus,
$t$-$J$ model is numerically easier than the Hubbard model in DMRG
approach.
\end{references}
\end{document}